\documentclass[conference]{IEEEtran}

\IEEEoverridecommandlockouts
\IEEEpubid{\makebox[\columnwidth]{ 979-8-3503-5067-8/24/\$31.00~\copyright2024 IEEE \hfill} 
\hspace{\columnsep}\makebox[\columnwidth]{ }}
\IEEEpubidadjcol

\usepackage{cite}
\usepackage{amsmath,amssymb,amsfonts}
\usepackage{algorithmic}
\usepackage{graphicx}
\usepackage{textcomp}
\usepackage{xcolor}
\usepackage{romannum}
\usepackage{multirow}

\def\BibTeX{{\rm B\kern-.05em{\sc i\kern-.025em b}\kern-.08em
    T\kern-.1667em\lower.7ex\hbox{E}\kern-.125emX}}

\DeclareRobustCommand{\IEEEauthorrefmark}[1]{\smash{\textsuperscript{\footnotesize #1}}}

\begin{document}

\title{FemQuest - An Interactive Multiplayer Game to Engage Girls in Programming}


\author{\IEEEauthorblockN{Michael Holly \IEEEauthorrefmark{\romannum{1}}, Lisa Habich \IEEEauthorrefmark{\romannum{2}}, Maria Seiser \IEEEauthorrefmark{\romannum{3}}, Florian Glawogger \IEEEauthorrefmark{\romannum{4}}, Kevin Innerebner \IEEEauthorrefmark{\romannum{5}}}
\IEEEauthorblockA{\textit{Graz University of Technology} \\
Graz, Austria \\
Email: \IEEEauthorrefmark{\romannum{1}}michael.holly@tugraz.at,
\IEEEauthorrefmark{\romannum{2}}lisa.habich@tugraz.at,
\IEEEauthorrefmark{\romannum{3}}seiser@tugraz.at,
\IEEEauthorrefmark{\romannum{4}}glawogger@tugraz.at,
\IEEEauthorrefmark{\romannum{5}}innerebner@tugraz.at
}
\and
\IEEEauthorblockN{Sandra Kupsa \IEEEauthorrefmark{\romannum{6}}, Philipp Einwallner \IEEEauthorrefmark{\romannum{7}}}
\IEEEauthorblockA{\textit{Jugend am Werk Steiermark} \\
Graz, Austria \\
Email: \IEEEauthorrefmark{\romannum{6}}sandra.kupsa@jaw.or.at,
\IEEEauthorrefmark{\romannum{7}}philipp.einwallner@jaw.or.at
}
\and
\IEEEauthorblockN{Johanna Pirker}
\IEEEauthorblockA{\textit{Ludwig-Maximilians-Universität München} \\
Munich, Germany \\
jpirker@iicm.edu
}
}

\maketitle

\begin{abstract}

In recent decades, computer science (CS) has undergone remarkable growth and diversification. Creating attractive, social, or hands-on games has already been identified as a possible approach to get teenagers and young adults interested in CS. However, overcoming the global gap between the interest and participation of men and women in CS is still a worldwide problem. To address this challenge, we present a multiplayer game that is used in a workshop setting to motivate girls to program through a 3D game environment. The paper aims to expand the educational landscape within computer science education by offering a motivating and engaging platform for young women to explore programming quests in a collaborative environment. The study involved 235 girls and 50 coaches for the workshop evaluation and a subset of 20 participants for an in-game analysis. In this paper, we explore the engagement in programming and assess the cognitive workload while playing and solving programming quests within the game, as well as the learning experience and the outcome. The results show that the positive outcomes of the workshop underscore the effectiveness of a game-based collaborative learning approach to get girls interested in computer science activities. The variety of solutions found for the different tasks demonstrates the creativity and problem-solving skills of the participants and underlines the effectiveness of the workshop in promoting critical thinking and computational skills.

\end{abstract}


 \section{Introduction}

The gender gap in computer science persists to this day. With 25\% of the US-American \cite{statista20} and 32.8\% of the European workforce being female\cite{eurostat23} in computer-related and high-tech occupations, it remains a male-dominated field. The reasons why women aren't entertaining a career in science, technology, engineering, and mathematics (STEM) are multi-faceted but are often rooted in early childhood \cite{AMEMIYA2024105740}. Girls who are performing well in science-related subjects are often also doing well in non-science-related courses \cite{02bd2b68-en}. So even if they were likely to be successful in computer science or related careers, young girls are frequently influenced by gender stereotypes and unconscious biases to pursue other interests. Improving the attitude of girls towards programming remains a challenge, especially because interest in these subjects drops early on \cite{happe2021effective}. To achieve a higher number of women working and being educated in STEM fields, girls' motivation needs to be encouraged and their self-confidence strengthened to approach these subjects. A diverse workforce offers the opportunity to develop better products for a broader spectrum of consumers and better understand their needs. More women working in science and technology promises to increase social equity and economic growth. In recent years, innovative educational strategies, such as gamification and virtual learning environments, have shown promise in engaging learners. Gamification can successfully enhance learners' motivation and engagement, offering an effective alternative to traditional learning methods \cite{papastergiou2009digital, partovi2019effect}. Virtual worlds provide diverse opportunities to engage learners through communication and social interaction \cite{Moschini2010}. Additionally, competitive and collaborative approaches have been shown to enhance learning outcomes in computer science education \cite{cerny2011competitive}. Adjusting these concepts and methods for females can help to overcome the gender gap in this field.

In previous studies, we have already investigated different motivating and engaging learning methods for STEM education. In a virtual environment, we applied a game-based learning approach in combination with chess puzzles to engage students in learning computer science concepts \cite{holly2022interactive}. Using a more immersive environment, we evaluated the engagement and learning experience from the students' and teachers' perspectives and potential use cases for school classes \cite{holly2021designing}. In this paper, we want to focus on girls and young women, introducing FemQuest - an interactive multiplayer game designed to engage players in programming. By assessing engagement, workload, and the learning experience, we aim to provide insights into the effectiveness of the learning tool for fostering interest and proficiency in programming among female learners.
The main research objectives are defined as:
\begin{itemize}
    \item Investigating the engagement of girls in introducing programming within a story-based multiplayer game in a workshop setting. 
    \item Assessing the girls' cognitive workload while playing and solving different programming activities.
    \item Examining the learning experience and the outcome after completing the team play experience.
\end{itemize}

\paragraph*{Contribution}
This paper presents a workshop setting involving 235 girls and 50 coaches, evaluating the feasibility of using a multiplayer game to engage girls in programming. A subset of 20 participants was selected to investigate the learning experience, focusing on engagement, workload, and learning outcomes using in-game measurements.  
\section{Background and Related Work}

In general, girls and women are often less interested in computer science (CS) than men. Even those who are interested are less likely to see it as a possible career path \cite{vrieler2021computerscienceclub}. Promoting the motivation and engagement of girls across STEM subjects to overcome the prevailing gender gap in these fields has been the focus of several studies in recent decades \cite{de2020addressing, milgram2011recruit, burns2016girls, sharma2021improving}. 
Several strategies such as creating attractive, social, or hands-on games have already been identified as possible approaches to get girls and young women interested in CS \cite{happe2021effective}. De Carvalho et al. \cite{de2020addressing} introduced the CODING4GIRLS initiative, which is dedicated to teaching girls coding using a game design and development process based on design thinking methodology that emphasizes creativity and human-centered solutions. The use of games to promote educational content has proven to be a successful strategy to improve learners' motivation \cite{papastergiou2009digital}. Compared to traditional learning methods, game-based digital learning offers a powerful way to not only improve the attitude of students toward increasing their performance but also improve their creativity when it comes to problem-solving \cite{partovi2019effect}. Furthermore, virtual worlds offer a powerful platform to engage learners in educational settings through communication and social interaction \cite{Moschini2010}. Gütl \cite{Guetl2011} suggests that such digital spaces have the potential to address collaboration issues prevalent in existing technologies. These environments offer a variety of tools for interacting and socializing that enable groups to work effectively together in different learning scenarios. In combination with verbal and non-verbal communication and creative abilities, avatars are important elements for promoting effective learning in groups \cite{franceschi2008engaging}. Furthermore, Crellin et al. \cite{crellin2009virtual} demonstrated the versatility of virtual worlds in various CS domains, serving as development environments, collaboration platforms, or simulation environments. Cerny and Mannova \cite{cerny2011competitive} further showcased a competitive and collaborative approach to improve learning outcomes in computer science education.
Visual programming environments with a block-based system are widely used to teach users programming concepts and computational thinking \cite{weintrop2019block}. Lin and Weintrop \cite{lin2021landscape} analyzed 46 block-based programming environments and explored different design approaches to promote the transition to text-based programming, such as blocks-only, dual-modality, one-way transition, and hybrid. Scratch allows users to learn programming by creating interactive stories, games, and animations using a visual programming language \cite{maloney2010scratch}. The tool Snap! extends this approach by enabling the creation of custom blocks and control structures, enhancing its suitability for more advanced introductions to computer science \cite{RomagosaiCarrasquer2019snap}. The block-based programming environment \textit{Alice} facilitates the simple creation of animations, interactive stories, and simple 3D games through creative exploration. It has shown that this has a positive effect on performance and retention of the learning content \cite{cooper2003alice, cooper2003using}. As another tool, Greenfoot specializes in the development of interactive graphical applications allowing students to quickly and easily create engaging programs while learning fundamental programming concepts \cite{kolling2010greenfoot}. All these tools possess a visual nature, aim to generate engagement through exciting activities, and focus on introducing programming to undergraduate students.
Dele-Ajayi et al. \cite{dele2015girls} discover engagement factors such as the level of challenge, the clarity of the goal, social interaction, immersion, and feedback that influence girls' interests and determine the characteristics of the game that are essential to engage them through games. The literature review by Hong et al. \cite{hong2024approaches} classifies game elements for educational gamification into five categories: performance, ecological, social, personal, and fictional. These categories distinguish the various elements used in educational settings. Gamification elements can be classified as intrinsic or extrinsic based on their purpose. Performance elements are extrinsic and provide feedback, while social and personal elements are intrinsic, engaging learners through actions and peer interactions. Ecological and fictional elements combine both extrinsic and intrinsic aspects, creating interactions through environmental contexts.
By embedding programming tasks within a story, girls can receive a higher motivation to learn programming principles. 
Kelleher \cite{kelleher2008using} demonstrated that the focus on storytelling makes learning programming more attractive to girls.
Previous research has shown that certain game characteristics and elements are favorable for girls to be engaged when playing computer games. Female players tend to prefer non-violence and non-competitive game genres compared to male ones and dislike overly sexualized in-game characters \cite{hartmann2006gender, vermeulen2011girls}. For games emphasizing social interaction, research indicates that girls generally prefer games with social features \cite{hartmann2006gender}.
An approach to increase the engagement of girls in computer science is to teach them in single-gender classes. Marquardt et al. \cite{marquardt2023singlegenderclasses} showed that they are significantly more open to CS-related topics and that the intervention evoked significantly more positive reactions than in girls from mixed-gender classes. In the following section, we describe a multiplayer workshop setting in a single-gender class where girls play together to solve programming tasks in a 3D game environment.
\section{Multiplayer Game Setting}

\begin{figure}[!ht]
    \centerline{\includegraphics[width=0.47\textwidth]{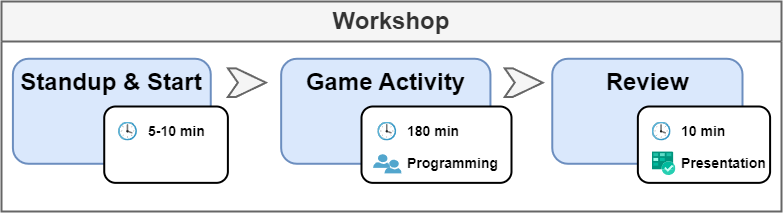}}
    \caption{Workshop Structure and Activities}
    \label{fig:WsStructure}
\end{figure}

FemQuest is designed as a multiplayer experience to help girls overcome hesitations about computer science, especially programming, and to encourage them to choose a career in this field. It is implemented in Unity\footnote{https://unity.com/} and supports mobile devices and a desktop application. To design the gaming application according to the target group's interests, girls and young women were involved in the development process. Their feedback was integrated according to the evolutionary development process \cite{Sommerville2007}. In a four-hour workshop setting, girls have to solve different programming tasks in teams, applying CS concepts such as conditions, loops, and variables using a block-based programming interface integrated into the game. The workshop is structured in three phases: (1) stand-up \& start, (2) game activity, and (3) review. Fig. \ref{fig:WsStructure} illustrates the workshop setting and the activities during the workshop. At the beginning of the workshop, all participants get together in a stand-up meeting to start as a group of 5-6 people. Afterward, they join the multiplayer session hosted by the workshop leader and play the game.
In this game session, they must collect items to unlock quests and solve programming puzzles. Each user solves the programming tasks individually and collaborates with the other players by unlocking quests to progress in the game level.
The host can intervene and define how many players must solve a specific quest. The players have to fulfill this setting to complete the level. Finally, in the review phase, the host leader discusses the solutions with the participants and asks them for feedback.

\subsection{Game Story}

\begin{figure}[!ht]
    \centerline{\includegraphics[width=0.47\textwidth]{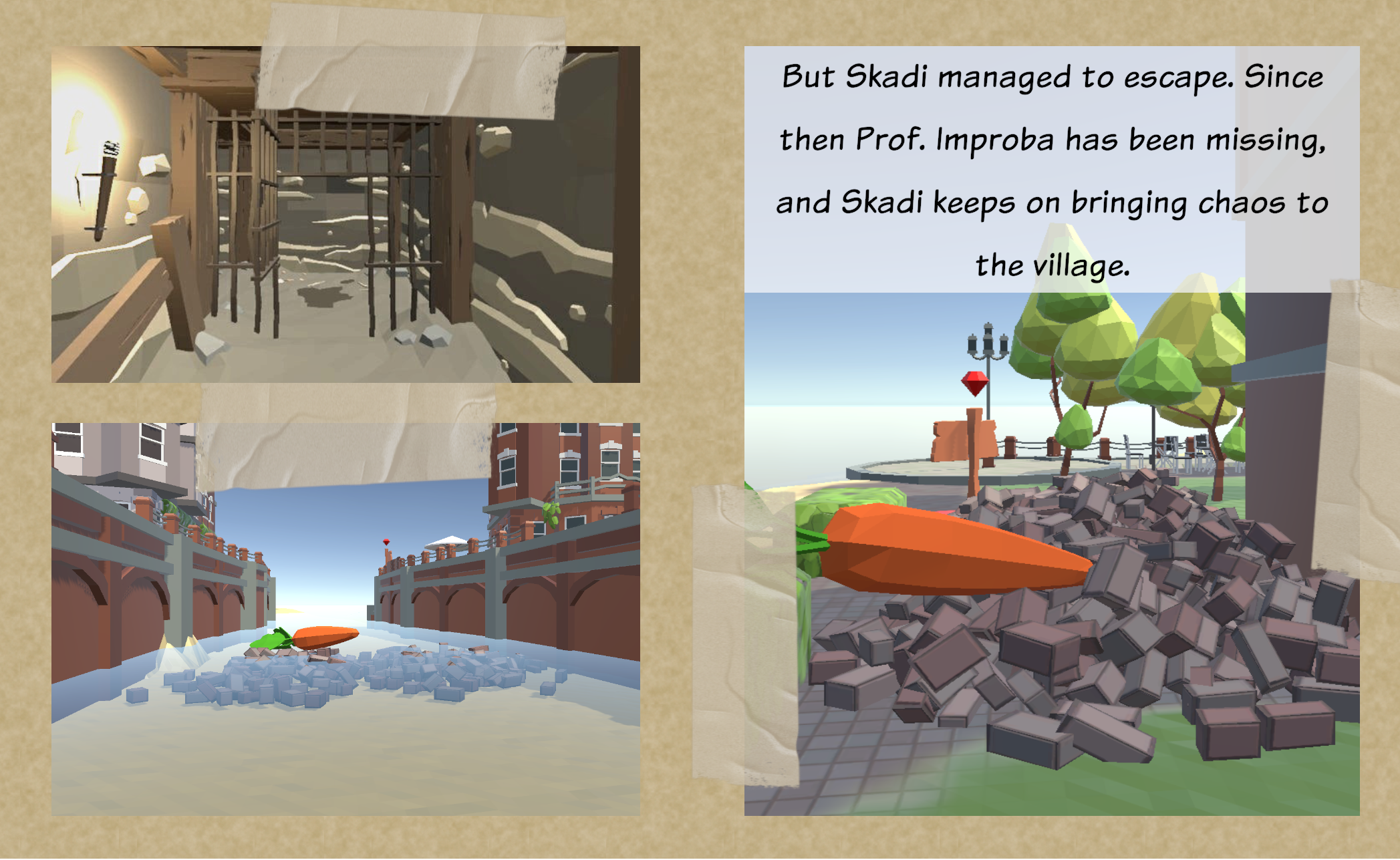}}
    \caption{Game story presented in a comic sequence.}
    \label{fig:comic}
\end{figure}

The game consists of three levels, where each level starts with a comic that introduces the story and the objectives of the level. Fig \ref{fig:comic} shows an exemplary clip of the introduction comic explaining the storyline. The story is about a clever fox created by a professor who seeks aid to rescue a village from a creature causing chaos. As a consequence, buildings are destroyed, the power supply has been disrupted, and the professor has been captured. Along the way, they must reconstruct buildings, restore power supply, gather ingredients for an effective antidote, and finally compete against the antagonist in a showdown. Throughout the game, players receive guidance and assistance and interact with the in-game characters. With each challenge overcome, players edge closer to restoring harmony and vanquishing the threat looming over the village.

\subsection{Game View}

\begin{figure}[!ht]
    \centerline{\includegraphics[width=0.47\textwidth]{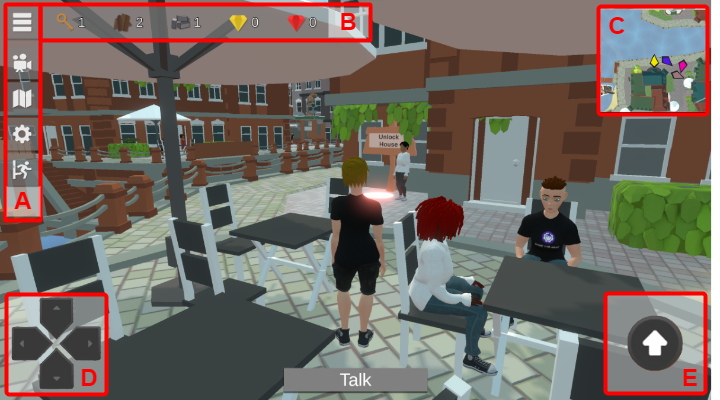}}
    \caption{Game View: (A) Menu, (B) Inventory Bar, (C) Mini-Map, (D) Move-Controls, (E) Jump}
    \label{fig:gameView}
\end{figure}

The game view includes the menu and inventory bar, the mini-map, and the move controls, as well as the player view (see Fig. \ref{fig:gameView}). Via the menu, the player can access the camera settings (first-person or third-person), the mini-map, sound, and language settings, and can switch the controls between joystick and d-pad. The players are represented by a customizable avatar, which the player can change before starting the level. Users can navigate through the virtual 3d environment via controls on the screen or a keyboard and can orient themselves using the mini-map. In this view, all quests are marked as red dots and players as colored arrows. For a better overview of the level, the users can expand the map by clicking on it. In the level, players must collect items and accept different quests. Some of the collectibles are individual for each player, and some are shared between the players so that they have to collaborate. Non-player characters (NPCs) in the level provide hints and tips for the player. The programming tasks are indicated with a red quest marker and an NPC next to it, explaining the mission and the items that are required to accept the challenge. When the player accepts, the game loads into the programming view, where the quest has to be solved. Once the quest has been completed, the player's inventory gets reduced accordingly and the player receives a reward for solving the quest. Furthermore, the red marker on the map disappears.

\subsection{Programming View}

\begin{figure}[!ht]
    \centerline{\includegraphics[width=0.47\textwidth]{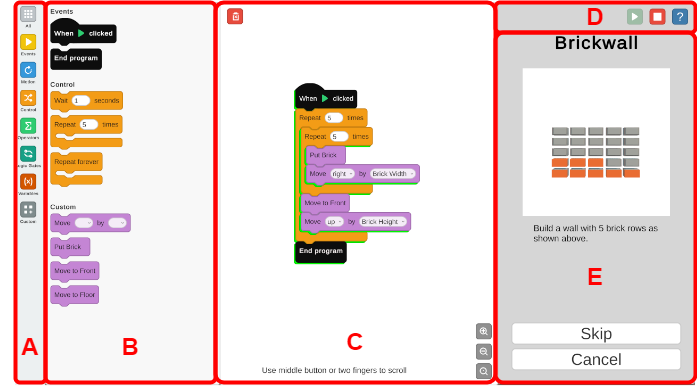}}
    \caption{Programming View: (A): Block Categories, (B) Blocks, (C) Programming Area, (D) Execution Bar, (E) Quest View}
    \label{fig:progView}
\end{figure}

The programming view is based on the Unity Block Engine\footnote{https://meadowgames.com/} asset which allows players to code within the game. It is divided into three main parts: (1) programming blocks, (2) programming area, and (3) quest view. Fig. \ref{fig:progView} gives an overview of the different panels and elements. The right panel presents the quest that has to be solved via block programming. By adding blocks from the block panel to the programming area, users can define a command sequence that is executed when the run button is pressed. Users can observe the outcome in the quest view and get visual feedback on whether the quest has been completed (green) or not (red). To assist players in solving the tasks, the app provides hints and tips with detailed explanations by clicking on the help button. After completing all mandatory sub-quests, the player returns to the game view to continue with the level. The player also has the option to cancel a quest at any time and resume it later.

\section{Study Design}

The study aimed to evaluate girl's engagement level in programming when playing a multi-user game within a workshop setting. The evaluation included a workshop evaluation and an in-game assessment. The first experiment evaluated the workshop setting, and the second experiment investigated the learning experience through additional in-game measurements using a subset of participants.

\subsection{Material and Setup}

To recruit participants, girls, and coaches from a job orientation course offered by a local social service provider were invited to participate in the study. Each participant was provided with an Android 13 tablet with a 10.5” display running the game application. One device was used for hosting the game session via a local WiFi network connection. To join the game session, the participants were asked to enter the host IP address in the game setting. To load the survey and upload data within the game, we used the LimeSurvey RemoteControl 2 API \cite{LimeSurvey2022}. 

\subsection{Workshop Evaluation}
For the workshop evaluation, we performed 40 workshops with girls and ten additional workshops focusing on coaches. The workshop itself was conducted as described before. At the beginning, the workshop leader gave a short introduction of the overall procedure, the game story, and the different tasks. Afterward, the participants had to play the game for three hours to complete at least one level. During this stage, the workshop leader was available to step in and give support. At the end of the workshop, they discussed the solutions in the group and were asked to give feedback. We asked them five feedback questions on a Likert scale between 1 (strongly disagree) and 5 (strongly agree) to assess their satisfaction and one open question to report optional feedback.

\subsection{In-Game Assessment}
To extend the study, we selected a representative subset of 20 participants. After the stand-up \& start meeting, the participants were asked to fill out a pre-questionnaire about their age, level of education, and previous experience with e-learning tools. In the game session, the participants played the first level of the game where they had to perform the following programming quests in ascending difficulty including three optional quests:
\begin{enumerate}
    \item \textit{Fence} quests where the users have to repair, close, and open a fence using a simple block command.
    \item \textit{House} and \textit{Bridge} quests where users must rebuild a house and a bridge by combining block commands with loops.
    \item \textit{Unlock} quests where users have to unlock a house to reach the next level using if and only if conditions. 
\end{enumerate}
While playing the game level, we measured the time per quest, the quest attempts, the number of collected items, and the programming solutions. The measurements were integrated into the game environment and started automatically when the user entered a level. After completion, participants were asked a post-questionnaire that included two standardized questionnaires to measure engagement, learning experience, and workload. The Web Based Learning Tools (WBLT)\cite{WBLT} questionnaire assesses learning, design, and engagement on a Likert scale from 1 (strongly disagree) to 5 (strongly agree). To assess the cognitive workload and task efficiency, we used the NASA Task Load Index (NASA-TLX) \cite{NASA-TLX}, whereby participants provided ratings of their workload using a Likert scale ranging from 1 (low) to 10 (high).

\subsection{Participants}
In total, 235 girls aged between 15 and 23 (AVG = 18.30, SD = 0.80) participated in the workshop evaluation. The participants represent a diverse group of girls and young women from different social and cultural backgrounds confronted with challenges such as unemployment, educational dropout, mental health problems, or social exclusion. Additionally, 50 job coaches with experience supporting and guiding young people in their career choices were asked to participate in the study. For the additional in-game assessment, we selected a subset of 20 participants considering their experience with computer usage, video games, and programming, as well as their educational level, social and cultural background. One of the girls had a primary education level, five had completed lower secondary school, eight had upper secondary school, and six had other qualifications. In the pre-questionnaire, we asked them to rate their experience with computers, video games, and programming on a Likert scale from 1 (low) to 5 (high). They rated their experience with computers as moderate (AVG=3.60, SD=1.10), video games slightly higher (AVG=3.85, SD=1.04), and programming as low (AVG=2.40, SD=1.47). Seven of them had already used an e-learning tool before.
\section{Results}

This section presents the results of the workshop evaluation and the in-game assessment with a focus on the learning experience, engagement, and workload. The findings are categorized into general feedback from the workshop, questionnaire data, and measurements recorded during gameplay.

\subsection{Workshop Feedback}
To obtain feedback on the workshop setting, we asked both girls and coaches to rate their overall satisfaction on a Likert scale between 1 (strongly disagree) and 5 (strongly agree). Table \ref{tab:WS_Feedback} shows the results of the different statements of both groups. Overall, the feedback from all participants was generally positive. Both groups stated that they were satisfied with the content presented in the workshop and agreed with the statement that the workshop met their expectations, with the coaches giving a slightly higher rating. Furthermore, knowledge transfer was rated as clear, and participants were inclined to recommend the workshop setting to others. The coaches found the setting and materials valuable and would use them. In the open questions, the girls reported that they found the workshop interesting and enjoyable. They gave positive remarks on the instructions by the workshop leader and found that the way the instructions were presented made the session enjoyable. A few mentioned that they learned new things while having fun playing the game. However, some participants found the programming aspect challenging and suggested a more beginner-friendly explanation. In addition, the coaches reported the overall positive atmosphere and the well-structured content. Nevertheless, some coaches felt that certain content was unnecessarily complex for instructional purposes and more suitable for school settings. 

\begin{table}[!ht]
\centering
\caption{Results of the workshop feedback on a Likert Scale between 1 (Strongly disagree) and 5 (Strongly agree)}
\label{tab:WS_Feedback}
\begin{tabular}{p{4.5cm}cccc}
\hline
                                                               & \multicolumn{2}{c}{Girls} & \multicolumn{2}{c}{Coaches} \\
                                                               & AVG         & SD          & AVG           & SD           \\ \hline
I was satisfied with the workshop.                             & 4.21        & 0.83        & 4.58          & 0.68         \\
The content of the workshop was fully covered.                 & 4.27        & 0.92        & 4.61          & 0.61         \\
The method of knowledge transfer was clear to me.              & 4.24        & 0.90        & 4.45          & 0.71         \\
The workshop met my expectations.                              & 4.18        & 1,06        & 4.48          & 0.76         \\
I would recommend the workshop.                                & 4.13        & 1.08        & 4.36          & 0.93         \\ 
I can use the materials given in my daily work.             & -           & -           & 4.11          & 1.06         \\
The materials given are suitable for use in my daily work.     & -           & -           & 4.37          & 0.94         \\ \hline
\end{tabular}
\end{table}

\subsection{Learning, Design, and Engagement}

To evaluate the learning experience and the effectiveness, we asked the participants to rate the WBLT categories learning, engagement, and design on a Likert scale from 1 (strongly disagree) to 5 (strongly agree) - see Table \ref{tab:WBLT}. The results show positive feedback across all three categories - Learning (AVG=3.53, SD=0.99), Design (AVG=3.36, SD=0.89), Engagement (AVG=3.51, SD=0.96). They described the game as conducive to learning and rated elements such as the use of learning objects and the effectiveness of the feedback as positive. While graphics and animations received a slightly lower rating, the game successfully taught new concepts, and participants perceived it overall as beneficial for learning. In terms of design, participants appreciated the utility of help features and found that instructions were generally easy to follow, although there is room for improvement in clarity. The usability and organization of the game were evaluated as above average. Regarding engagement, participants enjoyed the game theme and perceived it as making learning enjoyable. Participants also rated that they would use the game again.

\begin{table}[!ht]
\centering
\caption{Results of the WBLT on a Likert Scale between 1 (Strongly disagree) and 5 (Strongly agree)}
\label{tab:WBLT}
\begin{tabular}{p{1.2cm} p{4.8cm} cc}
\hline
                            &                                                   & AVG  & SD   \\ \hline
\multirow{5}{*}{Learning}   & Working with the learning object helped me learn. & 3.60 & 0.88 \\
                            & The feedback from the game helped me learn.       & 3.50 & 0.83 \\
                            & The graphics and animations helped me learn.      & 3.20 & 1.10 \\
                            & The game helped teach me a new concept.           & 3.80 & 1.15 \\
                            & Overall, the game helped me learn.                & 3.55 & 1.00 \\ \hline
\multirow{4}{*}{Design}     & The help features were useful.        & 3.50 & 0.69 \\
                            & The instructions were easy to follow. & 3.35 & 0.93 \\
                            & The game was easy to use.                         & 3.30 & 0.98 \\
                            & The game was well organized.                      & 3.30 & 0.98 \\ \hline
\multirow{4}{*}{Engagement} & I liked the overall theme of the game.            & 3.65 & 0.93 \\
                            & I found the game engaging.                        & 3.25 & 0.85 \\
                            & The game made learning fun.                       & 3.60 & 1.00 \\
                            & I would like to use the game again.               & 3.55 & 1.05 \\ \hline
\end{tabular}
\end{table}

\subsection{Workload Experience}

To assess the overall performance, effort, and mental and physical demand, we used the NASA TLX. We asked the participants to rate their experience on a Likert scale between 1 (low) and 10 (high). The results, summarized in Table \ref{tab:Workload}, provide information on the different dimensions of the workload during the game. Overall, participants reported moderate mental, physical, and time demands, suggesting that the tasks required appropriate cognitive processing and that dealing with time restrictions was not perceived as a significant constraint. Participants felt relatively confident in their performance with a moderate level of effort. Nevertheless, there were reported experiences of frustration, reflecting that some aspects of the task were frustrating or challenging, underscoring the potential for improvement in the task design.

\begin{table}[!ht] 
\centering
\caption{Results of the NASA-TLX on a Likert Scale between 1 (low) and 10 (high)}
\label{tab:Workload}
\begin{tabular}{lcc}
\hline
                & AVG  & SD   \\ \hline
Mental Demand   & 5.20 & 2.12 \\
Physical Demand & 5.20 & 1.74 \\
Temporal Demand & 4.60 & 2.19 \\
Performance     & 7.05 & 2.01 \\
Effort          & 6.01 & 2.17 \\
Frustration     & 5.35 & 1.93 \\ \hline
\end{tabular}
\end{table}

\subsection{Game Measurements}

\begin{figure}[ht]
    \centerline{\includegraphics[width=0.47\textwidth]{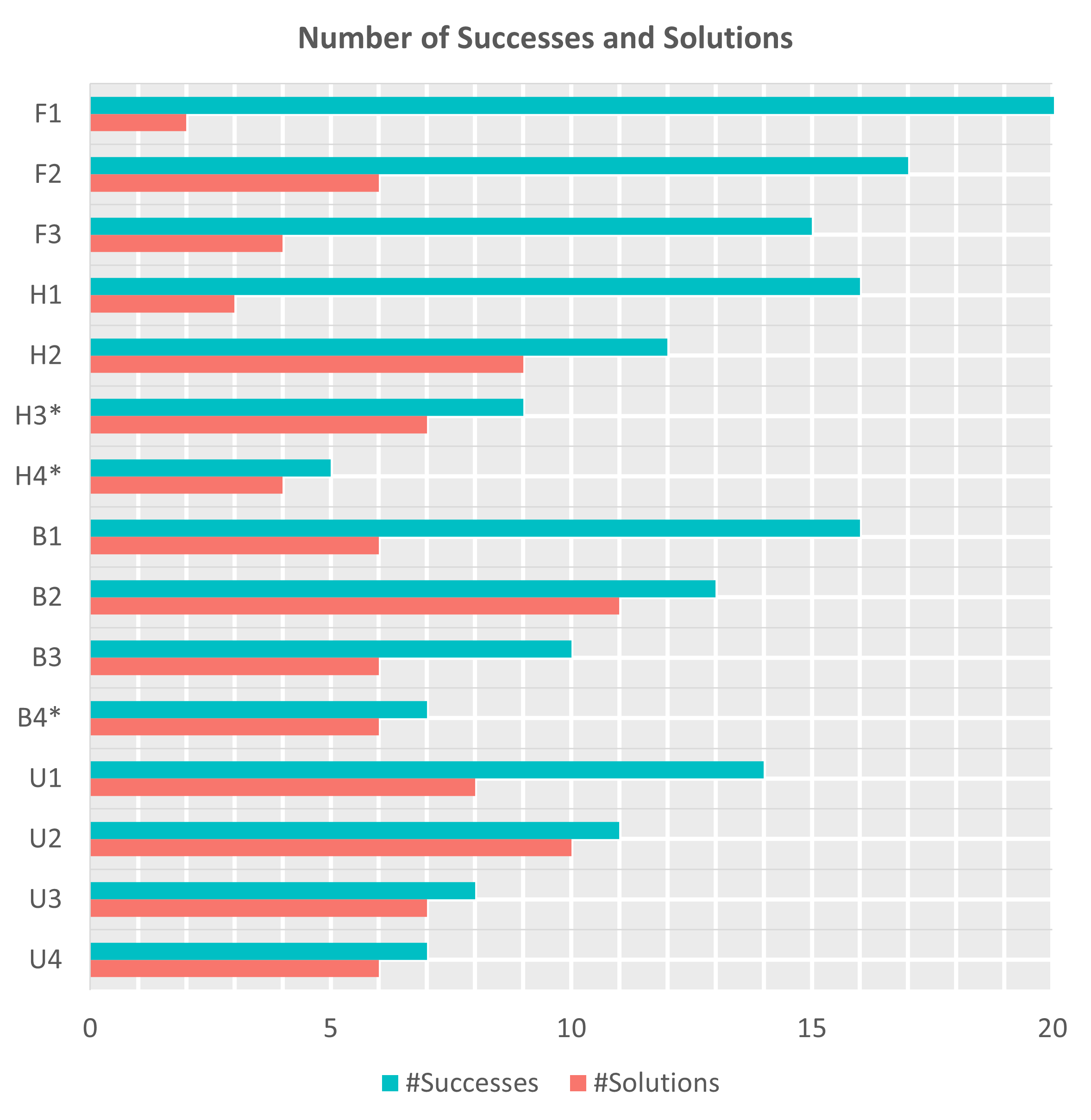}}
    \caption{Number of successes and solutions}
    \label{fig:solutions}
\end{figure}

During the game session, participants completed four main quests: Fence (F), House (H), Bridge (B), and Unlock (U).
To complete the level, all the mandatory quests had to be completed, while the three optional quests $Q*$ could be skipped by the players. To assess the gaming behavior, we recorded in-game measurements such as the collected items, quest attempts, time per quest, as well as the number of successes and different solutions. To start a quest, players had to collect certain items and receive a special award for optional quests. On average, each player collected six items and solved one optional quest. In solving the quests, the players were free to find their solutions. Fig. \ref{fig:solutions} shows the number of successes and the number of different solutions found for each quest. The initial quests (F1-F3) were completed by most of the participants. Although the more challenging tasks were solved less often, the number of different solutions increased for the more complex ones.


\begin{figure}[ht]
    \centerline{\includegraphics[width=0.47\textwidth]{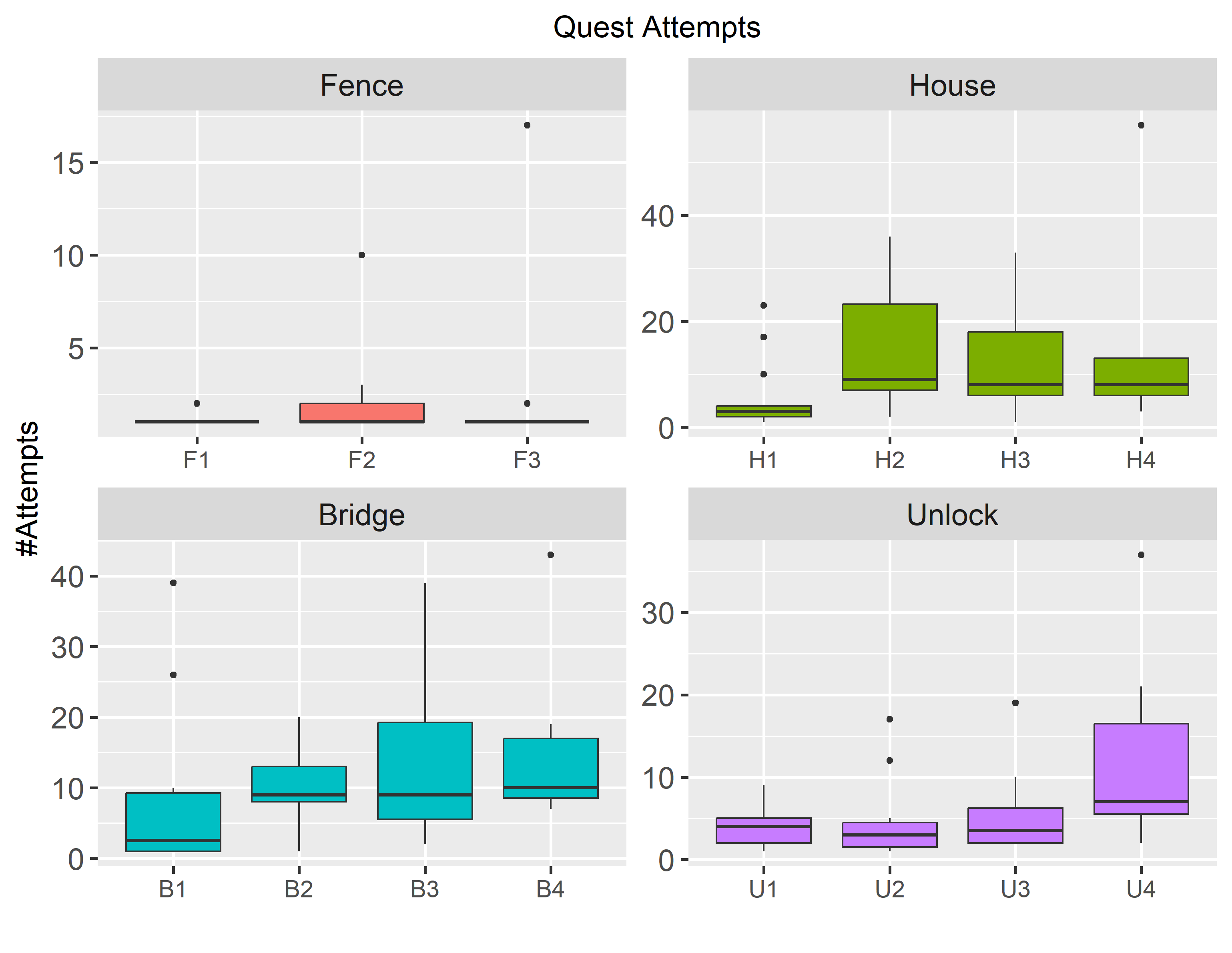}}
    \caption{Attempts per Quest}
    \label{fig:attempts}
\end{figure}

The average number of attempts required to complete a quest (see Fig. \ref{fig:attempts}) provides valuable insights into the difficulty or complexity participants encountered while attempting to solve the quests. Initially, the fence quests displayed a relatively low average number of attempts, suggesting a straightforward nature. On the contrary, the house and bridge quests demanded more attempts, indicative of their increased challenge level, requiring players to make more trial and error before succeeding. In the unlock quest, the average number of attempts has decreased for most of the sub-quests.

\begin{figure}[ht]
    \centerline{\includegraphics[width=0.47\textwidth]{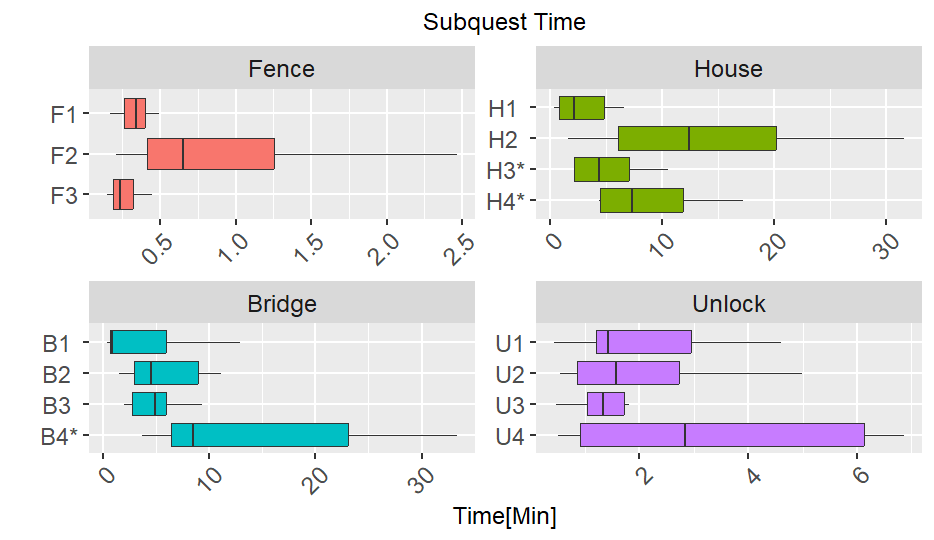}}
    \caption{Time per Subquest}
    \label{fig:time}
\end{figure}

Fig. \ref{fig:time} represents the mean duration, in minutes, of the participants required to complete each quest. Shows the efficiency and proficiency of participants in executing tasks within the game environment. Participants completed the initial fence quests on average in less than one minute. Other quests $Q$ and optional quests $Q*$ were required correspondingly longer ($Q$: AVG=4.68 min; $Q*$: AVG=9.39 min).

\section{Discussion}

The positive outcomes of the workshop underscore the effectiveness of employing a collaborative game-based approach to engage girls in computer science activities. The high level of satisfaction reported by both the girls and the coaches indicates that the workshop was an enjoyable and useful learning experience. However, feedback also shows potential for improvement, particularly in making certain aspects of programming tasks more accessible to participants. Addressing these challenges through clearer instructions and more beginner-friendly explanations could further enhance the workshop's effectiveness in catering to a diverse range of skill levels among participants.
The results of the learning, design, and engagement assessment further validate the efficacy of the game-based approach. Participants perceived the game as conducive to learning, with effective use of learning objects and feedback mechanisms contributing to their understanding of the concepts presented. Although suggestions for improvement in terms of clarity and organization indicate room for refinement, the overall enjoyment and engagement reported by participants underscore the value of incorporating gamification elements into educational activities. This is in line with the findings of Burns et al. \cite{burns2016girls} suggesting that playing or designing games can positively influence girls' perceptions of pursuing a career in CS. Crucially, these activities should emphasize personalization, collaboration, and the presence of female characters to enhance engagement and motivation among girls.
The NASA-TLX results provide insights into the workload experienced by participants during the game. While participants reported moderate mental, physical, and temporal demands, the presence of frustration suggests opportunities for optimizing task design to minimize cognitive load and enhance the user experience. By addressing these challenges, future iterations of similar workshops can strive to provide a more seamless and enjoyable learning experience for participants. Analysis of in-game measurements revealed interesting patterns in participants' gaming behavior, highlighting the progression of difficulty as participants advanced through the quests. The variety of solutions found for different quests demonstrates the creativity and problem-solving skills exhibited by participants, further emphasizing the effectiveness of the workshop in fostering critical thinking and computational skills.
The alignment of the workshop outcomes with previous studies on strategies to engage girls in computer science, such as the CODING4GIRLS initiative \cite{de2020addressing} and the benefits of single-gender classes \cite{marquardt2023singlegenderclasses}, provides valuable context for understanding the broader impact of gender-specific approaches in promoting girls' interest in STEM fields. 
Such initiatives are dedicated to teaching girls to program by using design thinking-based methods with a focus on creative and human-centered solutions. The variety of approaches participants employed to solve the different quests shows their creativity, adaptability, and willingness to explore alternative strategies within the game. The number of successes and different solutions provides insights into participants' problem-solving abilities, creative thinking, and adaptability within the game environment. The variability in solutions not only showcases participants' problem-solving skills but also underscores the flexibility and open-ended nature of the game environment, allowing for multiple avenues to achieve objectives. However, it should be noted that the variety of solutions can also be an indicator of potential issues where the task or instructions could be clarified or refined to reduce ambiguity and improve understanding. Quests with a higher number of programming blocks can tend to have a higher variety of solutions, highlighting the intricate relationship between task complexity and participants' problem-solving strategies. Analyzing the number of attempts provides valuable insights into the difficulty level of the quests, highlighting areas where participants may have faced challenges or required additional iterations to complete the quests. The house quest witnessed a decrease in the average number of attempts, indicating a potential adjustment in player strategies or a shift towards more efficient problem-solving approaches.
Additionally, insights from research on game-based learning environments \cite{happe2021effective}, visual programming environments \cite{weintrop2019block, lin2021landscape}, and educational gamification \cite{toda2019taxonomy} inform the theoretical underpinnings of the workshop's design and implementation, further validating its effectiveness in engaging and educating participants.
Overall, the workshop serves as a successful example of how innovative approaches, informed by research and tailored to the needs and preferences of participants, can effectively promote girls' interest and participation in CS activities. By leveraging the engaging nature of games and incorporating interactive and hands-on learning experiences, workshops like these could play a crucial role in inspiring the next generation of female computer scientists and closing the gender gap in STEM fields.

\subsection{Limitations}
The main limitation of the study is the relatively small size of the 20 participants in the subset for the detailed analysis. This fact may not fully capture the diverse range of all participants. Furthermore, the results include only information on short-term participation, workload, and learning outcomes, but the long-term effects and sustainability of the intervention are not addressed. A long-term study could provide a more comprehensive understanding of the impact of the game over time. Additionally, participants had different social and cultural backgrounds with a low level of attention, so short questionnaires in simple language had to be chosen. This could have a potential impact on the collected data and its ability to fully capture the complexity of the participants' experiences.
\section{Conclusion}

In conclusion, the findings from the evaluation highlight the effectiveness of the multiplayer workshop setting in engaging girls in computer science activities within a 3D game environment. The positive feedback received from both participants and coaches, along with the insights derived from in-game assessments, offer valuable guidance for refining and improving future iterations of similar workshops. Using the immersive and interactive nature of games and integrating hands-on programming tasks, such workshops have the potential to not only foster interest and engagement but also inspire and motivate girls to pursue further studies and careers in computer science.
Regarding learning experience and effectiveness, participants rated learning, engagement, and design positively. The game was seen as conducive to learning, with positive remarks on learning objects and feedback effectiveness. Although graphics and animations received slightly lower scores, participants found the game enjoyable and were willing to use it again. 
In general, the workshop and the game received positive feedback, suggesting the effectiveness of teaching new concepts, fostering engagement, and providing a pleasant learning experience. Possible improvements include clarifying instructions, simplifying complex content, and refining quest design to balance challenge and enjoyment.
We believe this approach can also be adapted for multi-gender classes and that further evaluations in school settings would provide insights into the suitability of the pedagogical model from both learners' and teachers' perspectives. This could involve gathering feedback on the game's accessibility, its effectiveness in promoting learning, and its integration into classroom instruction.

\section*{Acknowledgment}
This work was supported by the Federal Chancellery Republic of Austria.

\bibliographystyle{IEEEtran}
\bibliography{references}

\end{document}